# Solving the paradox of the folded falling chain by considering horizontal kinetic energy and link geometry

Hong-Hsi Lee[1,*], Chih-Fan Chen[2], I-Shing Hu[3]

1 Center for Biomedical Imaging, New York University School of Medicine, New York, NY 10016, USA

2 Department of Physics, University of California, Davis, CA 95616, USA

3 Department of Mathematics, National Taiwan University, Taipei, 10617, Taiwan

* Corresponding author: Hong-Hsi Lee, email: Hong-Hsi.Lee@nyumc.org

ORCID ID: https://orcid.org/0000-0002-3663-6559

## Abstract

A folded chain, with one end fixed at the ceiling and the other end released from the same elevation, is commonly modeled as an energy-conserving system in one-dimension. However, the analytical paradigms in previous literature is unsatisfying: The theoretical prediction of the tension at the fixed end becomes infinitely large when the free end reaches the bottom, contradicting to the experimental observations. Furthermore, the dependence of the total falling time on the link number demonstrated in numerical simulations is still unexplained. Here, considering the horizontal kinetic energy and the geometry of each link, we derived analytical solutions of the maximal tension as well as the total falling time, in agreement with simulation results and experimental data reported in previous studies. This theoretical perspective shows a simple representation of the complicated two-dimensional falling chain system and, in particular, specifies the signature of the chain properties.

## Keywords

Falling chain; horizontal kinetic energy; link number; link geometry

## 1 Introduction

The classic falling chain system consists of a folded chain with one end fixed on the ceiling and the other end released to fall from the same elevation (Calkin and March, 1989) (Figure 1a). The two ends of the chain are supposed to be close to each other, sometimes dubbed as "narrow catenary" scheme (Tomaszewski and Pieranski, 2005). To solve this problem, the chain is assumed to be energy-conserving, i.e. with no energy loss during the falling, and the kinetic energy of the system is solely contributed by the vertical component of the chain velocity over the free side (Calkin and March, 1989; Tomaszewski and Pieranski, 2005). In other words, the analytical paradigm is simplified as a one-dimensional model. Although the two-dimensional motion of the falling chain has been explored by using either numerical simulations (Schagerl et al., 1997; Tomaszewski and Pieranski, 2005) or a nonlinear FEM model (Orsino and Pesce, 2018), the analytical relations between the measurement (e.g., total falling time, maximal tension at the support) and the chain properties (e.g., chain number, link shape) are still unknown.

The tension at the fixed end/support, based on the theory in previous studies (Calkin and March, 1989; Tomaszewski and Pieranski, 2005; Wong et al., 2007), grows and becomes infinitely large when the free end reaches the bottom, obviously contradicting to the experimental observations (Calkin and March, 1989). This unrealistic theoretical prediction originates from the assumption of the link number $n \rightarrow \infty$ (Wong et al., 2007); to better estimate the maximal tension at the support, the motion of the last link of the free side has to be analyzed separately.

Here, we derive analytical solutions of physical quantities, such as the total falling time and the tension at the support, by considering (a) the horizontal kinetic energy throughout the falling process, (b) the geometry of each link, and (c) the particularity of the last link's motion. The derived physical quantities are highly related to the link number and the link geometry, indicating that the two parameters are the most representative chain properties. Furthermore, our analytical solutions are consistent with the numerical simulations in (Schagerl et al., 1997) and experimental data in (Calkin and March, 1989), demonstrating the applicability of our theory.

## 2 Theory: energy-conserving chain with horizontal kinetic energy

The conservation of mechanical energy is presumably satisfied for a falling chain system. In this case, the kinetic energy is usually considered exclusively from the vertical velocity component in previous studies (Calkin and March, 1989). However, when a link at the bottom transits from the free side to the fixed side, the link obtains a non-trivial horizontal velocity component. In other words, mechanical energy $E$ includes kinetic energies, $K_x$ and $K_y$, contributed by horizontal and vertical velocity components and the gravitational potential energy $U_g$. The law of conservation of energy then leads to

$$\frac{dE}{dt} = \frac{d}{dt}\left(K_x + K_y + U_g\right) = 0. \#(1)$$

It is straightforward to calculate $K_y$ and $U_g$ as in the standard models (Calkin and March, 1989):

$$K_y = \frac{1}{4}\mu(L - y)\dot{y}^2 \, , \#(2a)$$

$$U_g = -\frac{1}{4}\mu g(L^2 + 2Ly - y^2) \, , \#(2b)$$

where $\mu = M/L$ is the linear mass density, $M$ and $L$ are the total length and the total mass of the chain, $y$ is the distance traveled by the free end, and $g$ is the gravitational acceleration. To calculate the change rate of the horizontal kinetic energy $K_x$, we focus on the link at the bottom (turning area), where the transfer of $K_y$ to $K_x$ happens when the bottom link transits from the free side to the fixed side. Given that the bottom link has a length $l = L/n$, a mass $m = \mu l$, and a moment of inertia $I = C \cdot ml^2$ with respect to the pivot (point O in Figure 1a) with a geometry constant $C$ determined by the link's geometry, we calculate this link's vertical kinetic energy $\widetilde{K}_y$ and the corresponding change rate when this link is in an angle $\theta$ with respect to the horizontal direction, as defined in Figure 1a:

$$\frac{d\widetilde{K}_y}{dt} = -\frac{d}{dt}\left(\frac{1}{2}I \cdot \dot{\theta}^2 \cos^2\theta\right) = -\frac{1}{2}\mu l C \cdot \frac{d\dot{y}^2}{dt} \, , \#(3)$$

where the relation $\dot{y} = \dot{\theta} l \cos\theta$ is used. Assuming that the increase of the horizontal kinetic energy $K_x$ is contributed by the decrease of the $\widetilde{K}_y$, i.e. $\frac{dK_x}{dt} = -\frac{d\widetilde{K}_y}{dt}$, substituting Eq. (2) and Eq. (3) into Eq. (1), and using the initial conditions, $y(t = 0) = 0$ and $\dot{y}(t = 0) = 0$, we obtained

$$\dot{y}^2 = \frac{g(2Ly - y^2)}{(L-y) + \frac{2CL}{n}}\,,\#(4)$$

where the correction term in the denominator, $2CL/n$, leads to (a) a relationship between the total falling time $t_e$ and the link number $n$, Eq. (6), and (b) a reasonable estimate of the maximal tension at the fixed end of the chain (point A in Figure 1a), Eq. (10).

## 2.1 Falling time

By solving the differential equation in Eq. (4), the relationship of the falling time $t$ and the free side vertical displacement $y$ is given by

$$\sqrt{\frac{g}{L}}t = 2\sqrt{\frac{2C}{n}+1}\cdot E\left(\frac{1}{4}(\pi - 2\alpha)\middle|\frac{2}{\frac{2C}{n}+1}\right),\#(5)$$

where $\sin\alpha = 1 - y/L$, and $E(\cdot\mid\cdot)$ is the elliptic integral of the second kind. Then the total falling time $t_e$ is given by

$$t_e = t(y = L)\,,\#(6)$$

which explicitly relates $t_e$ to the link number $n$ and the link's geometry constant $C$.

## 2.2 Tension at the fixed end/support

The vertical component of the tension $T_y$ at the fixed end of the chain (point A in Figure 1a) is given by

$$T_y = Mg - \dot{P}_y\,,\#(7)$$

where $M = \mu L$ is the chain's total mass, and $P_y$ is the chain's total vertical momentum:

$$P_y = \mu\left(\frac{L-y}{2}\right)\dot{y}\,.\#(8)$$

Substituting Eq. (4) and Eq.(8) into Eq. (7), we obtained the tension at the fixed end:

$$\frac{T_y}{Mg} = 1 + \frac{1}{2L}\cdot\frac{(2Ly - y^2)}{(L-y)+\frac{2CL}{n}}\cdot\left[1 - \left(\frac{L-y}{2}\right)\cdot\left(\frac{-1}{2L-y} + \frac{1}{y} + \frac{1}{(L-y)+\frac{2CL}{n}}\right)\right].\#(9)$$

Eq. (9) is applicable only for $y \ll L - L/n$. To estimate the maximal tension $T_y(y = L)$, we consider the last link's transition from the free side to the fixed side as follows: When the free end of the last link reaches the bottom (Figure 1b), i.e. $y \to L$, the last link has a change of momentum along the vertical direction: $\delta\tilde{P} \simeq -\tilde{P}\delta\theta$, where $\tilde{P}$ is the last link's momentum, and $\delta\theta$ is the corresponding infinitesimal angular change (Figure 1b-c). Assuming that $\tilde{v}$ is the speed of the last link's free end as $y \to L$, we obtained $\tilde{P} = \frac{1}{2}m\tilde{v}$ and $\frac{\delta\theta}{\delta t} = \frac{\tilde{v}}{l}$. When the last link is in the horizontal position and is going to rotate toward the bottom, its free end has a velocity $\dot{y}(y = L - L/n)$. The speed of the free end $\tilde{v}$ hardly changes during the last link's rotation, since the gain of the gravitational energy from the last link is very small. By approximating $\tilde{v} \simeq \dot{y}(y = L - L/n)$, calculated by using Eq. (4), the maximal tension at the fixed end for $y = L$ is given by

$$T_y(y = L) = Mg - \frac{\delta\tilde{P}}{\delta t} = Mg + \frac{m}{2l}\tilde{v}^2$$

$$\simeq Mg + Mg\frac{n - \frac{1}{n}}{2(2C + 1)} . \#(10)$$

2.3 Kinetic energy

Finally, we calculated the horizontal and the vertical kinetic energies, $K_x$ and $K_y$, using Eq. (2a), Eq. (3) and Eq. (4):

$$\frac{K_x}{\frac{1}{4}MgL} = \frac{2C}{nL} \cdot \frac{(2Ly - y^2)}{(L - y) + \frac{2CL}{n}} , \#(11a)$$

$$\frac{K_y}{\frac{1}{4}MgL} = \frac{L - y}{L^2} \cdot \frac{(2Ly - y^2)}{(L - y) + \frac{2CL}{n}} , \#(11b)$$

and

$$\frac{K_x}{K_y} = \frac{2C}{n\left(1 - \frac{y}{L}\right)} . \#(11c)$$

Eq. (11) shows that the horizontal kinetic energy $K_x$ increases with the geometry constant $C$ and decreases with the link number $n$. When $C = 0$ or $n \to \infty$, $K_x \to 0$. The ratio of the horizontal and the vertical kinetic energies highly depends on $n$ and $C$, indicating that the two parameters are the most essential signature to describe the falling chain system.

It is worthwhile to notice that the vertical kinetic energy $K_y$ monotonically increases over time when $C = 0$, and increases in the beginning and decreases in the end when $C \neq 0$, with a maximum as $y \simeq L(1 - \sqrt[3]{C/n})$ for $n \gg 1$.

## 3 Methods

To validate the total falling time $t_e$ in Eq. (5) and Eq. (6), we compared our analytical solution with the simulation results in (Schagerl et al., 1997). The simulation of the falling chain in the fig. 12 of (Schagerl et al., 1997) was performed for $n$ = 11-321, and the moment of inertia (with respect to the pivot O in Figure 1a) for each oval link is approximated as the rectangular one's (fig. 2 in (Schagerl et al., 1997)):

$$I_1 \simeq 0.19\, ml^2 + \frac{1}{4} ml^2 \,, I_2 \simeq 0.26\, ml^2 + \frac{1}{4} ml^2 \,.$$

Here, we use the averaged moment of inertia to estimate the geometry constant $C \simeq 0.475$. Since the geometry constant $C$ and the link number $n$ are both known, we can estimate the simulated falling time in the fig. 12 of (Schagerl et al., 1997), based on our analytical solutions, Eq. (5) and Eq. (6).

Furthermore, we compared our solutions of tension $T_y$ at fixed end, Eq. (9) and Eq. (10), with the experimental data in the fig. 3 of (Calkin and March, 1989), where the normal linked chain was composed of $n$ = 81 links. Since the moment of inertia for each link was not given in (Calkin and March, 1989), we assumed that each link could be approximated by a homogenous rod with $C$ = 1/3. Then we estimate the experimental data of the tension $T_y$ in the fig. 3 of (Calkin and March, 1989), based on our analytical predictions, Eq. (9) and Eq. (10), with respect to the falling time $t$ in Eq. (5). The corresponding horizontal and vertical kinetic energies are also plotted with respect to the falling time $t$, based on Eq. (11).

## 4 Results

In Figure 2, the analytical solution of the total falling time $t_e$, based on Eq. (5) and Eq. (6), is consistent with the simulation results in the fig. 12 of (Schagerl et al., 1997). In contrast, the solution in previous studies without considering the geometry constant (Calkin and March, 1989), i.e. $C = 0$, fails to evaluate the dependence of the total falling time $t_e$ on the link number $n$.

In Figure 3, the analytical solution of the tension $T_y$ at the fixed end, based on Eq. (9), estimates the experimental data in the fig. 3 in (Calkin and March, 1989) equally well with or without considering the geometry constant $C$. However, the maximal tension $T_y(y = L)$ can only be estimated (≈ 25.3 $Mg$) by including a non-zero geometry constant $C$ in Eq. (10). The corresponding horizontal and vertical kinetic energies are shown in Figure 4, where the horizontal kinetic energy $K_x$ significantly increases when $t \to t_e$, i.e. the last few links reach the bottom.

## 5 Conclusions

Introducing a horizontal kinetic energy and a geometry constant for each link, we successfully evaluate the total falling time and the maximal tension at the fixed end for the classic folded falling chain problem. The analytical solutions are consistent with simulation results and experimental data in previous studies, indicating the applicability of our theory. The contribution of the horizontal kinetic energy is negligible only when the link number $n \to \infty$ or the geometry constant $C = 0$, neither of which is possible for an actual system in reality.

**Acknowledgement**

This is a pre-print of an article published in *Acta Mechanica*. The final authenticated version is available online at: https://doi.org/10.1007/s00707-018-2350-9.

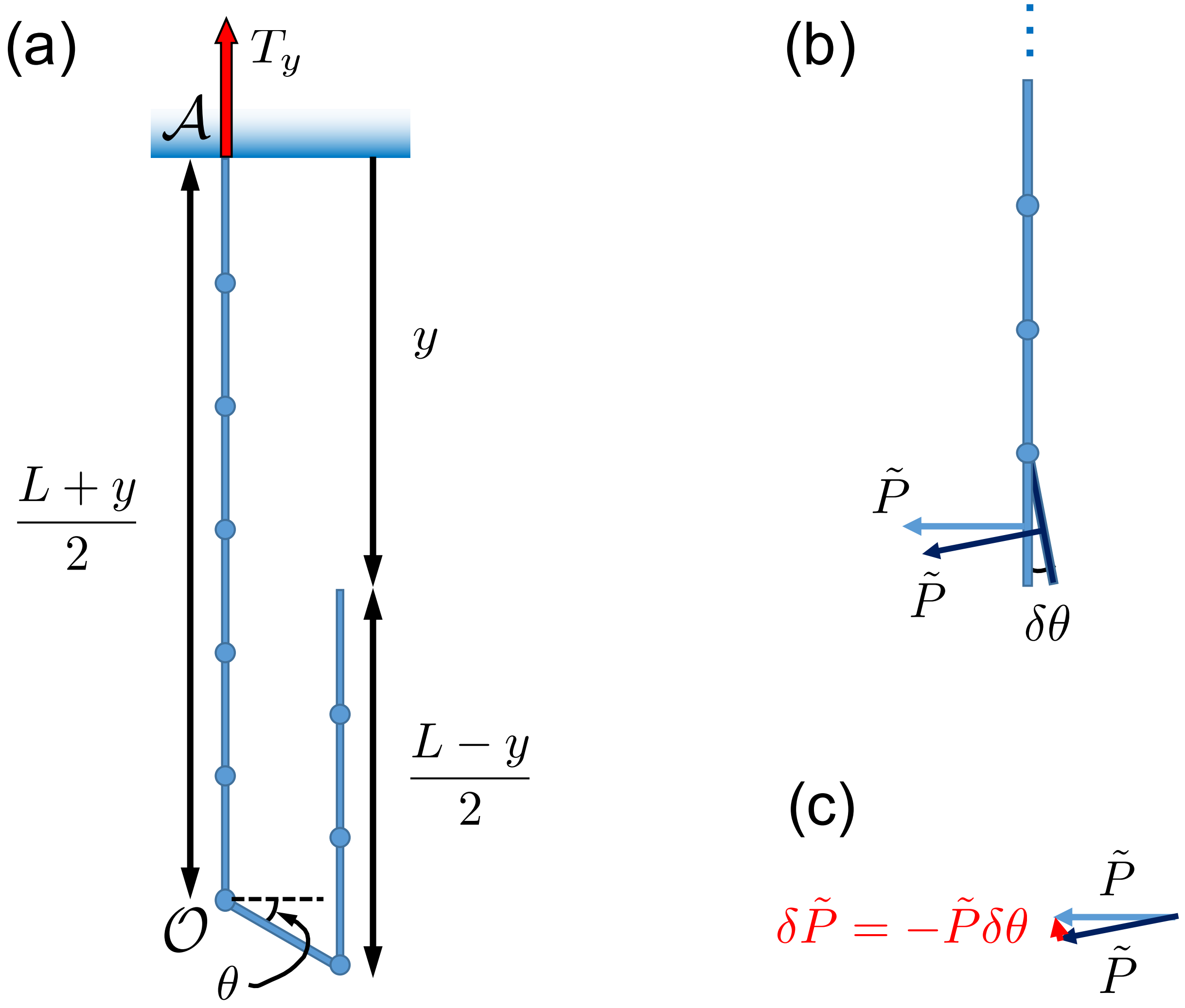

**Fig. 1**. (a) The folded chain with one end fixed at the ceiling (point A) and the other end released from the same elevation (Calkin and March, 1989). (b) When the last link's free end is going to reach the bottom, the last link changes the direction of its momentum $\tilde{P}$ with a small angle $\delta\theta$, and (c) the change of its momentum is $\delta\tilde{P}$.

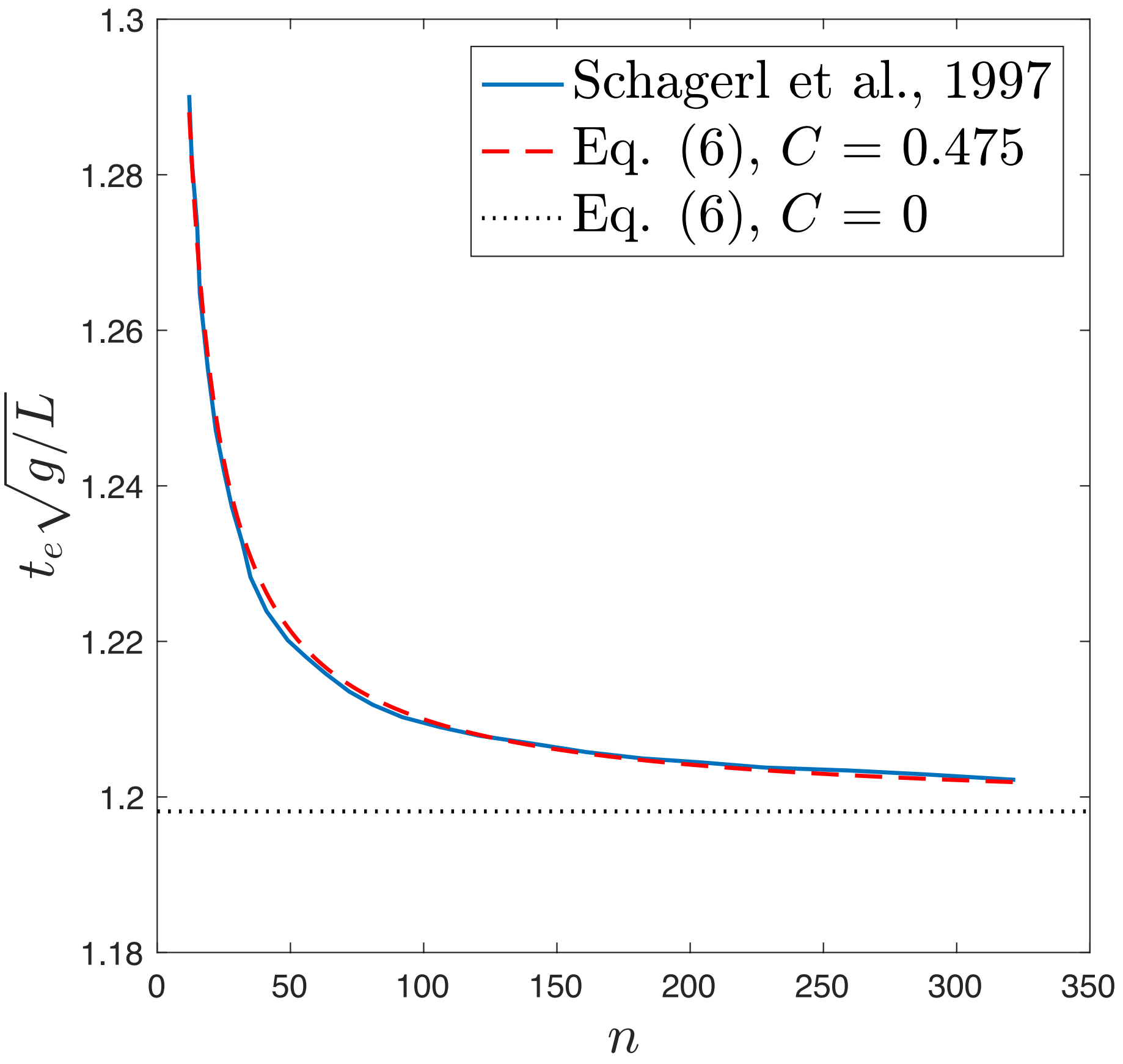

**Fig. 2**. The simulation results of the total falling time $t_e$ with respect to the link number $n$, cf. fig. 12 of (Schagerl et al., 1997) (blue solid line), is consistent with the analytical solutions, Eq. (5) and Eq. (6), when considering a non-zero geometry constant $C$ = 0.475 (red dashed line). The simulation result in this figure is adapted from (Schagerl et al., 1997).

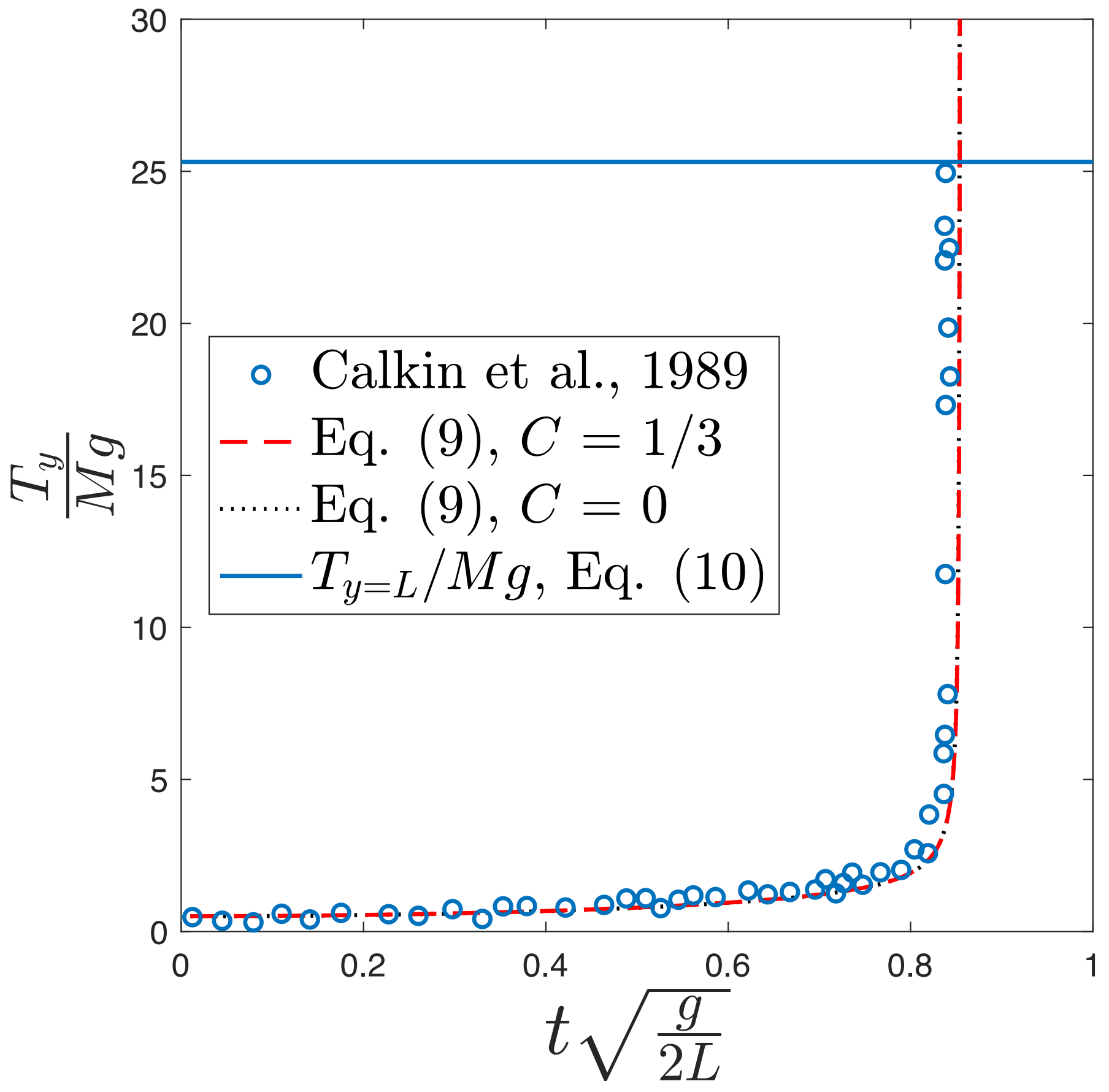

**Fig. 3**. The experimental data of the tension $T_y$ at the fixed end, cf. fig. 3 in (Calkin and March, 1989) (data point), is consistent with the analytical solution, Eq. (9), either with (red dashed line) or without (black dotted line) considering the geometry constant $C$. However, the maximal tension $T_y(y = L)$ can only be estimated by including a non-zero geometry constant $C$ = 1/3 in Eq. (10), yielding an estimate ≈ 25.3 $Mg$ (blue solid line). The data point in this figure is adapted from (Calkin and March, 1989).

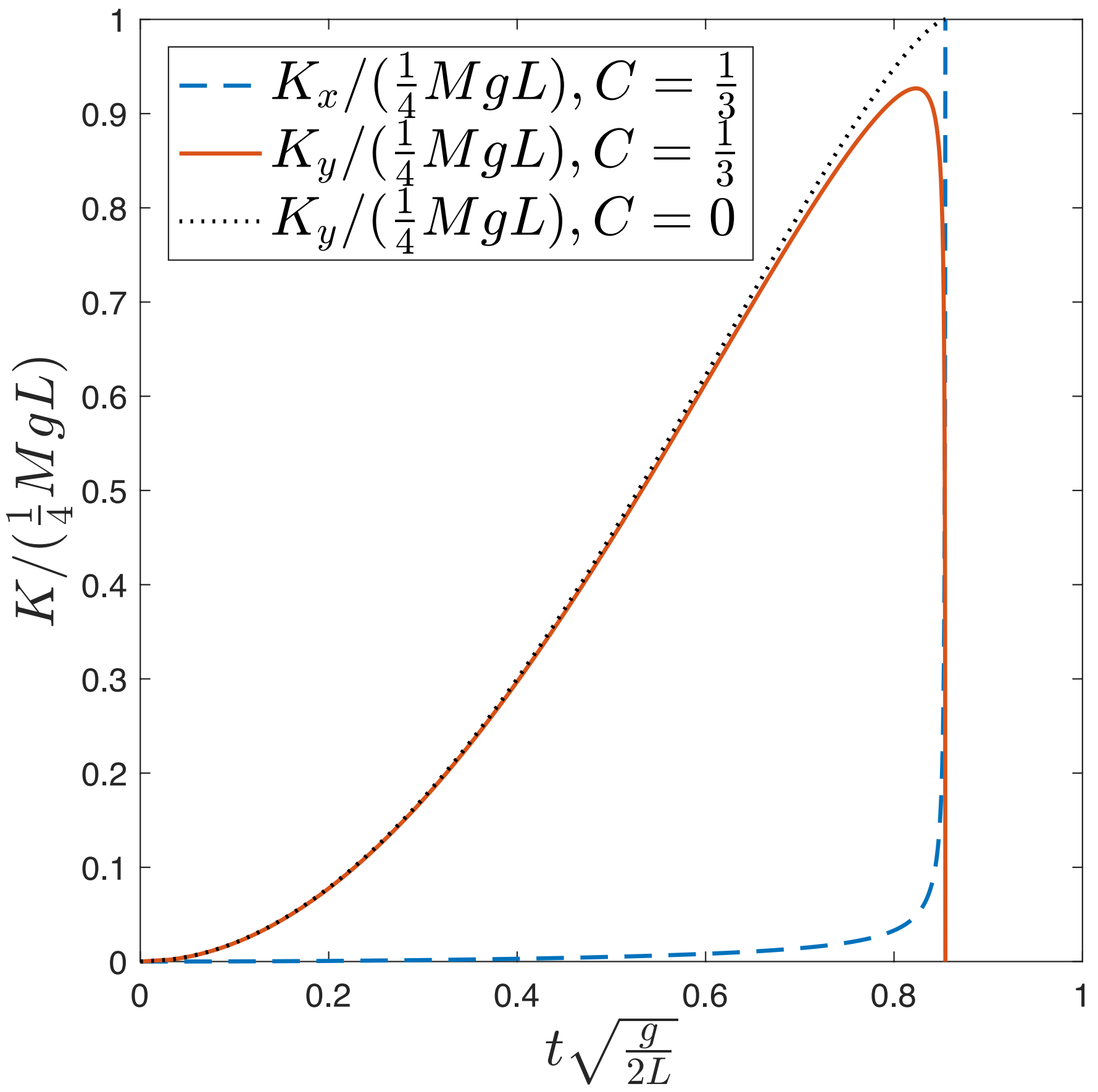

**Fig. 4**. The horizontal and the vertical kinetic energies, $K_x$ (blue dashed line) and $K_y$ (red solid line), with respect to the falling time $t$, based on Eq. (11). The link number $n$ = 81 and the geometry constant $C$ = 1/3 are chosen to match the experimental setup in (Calkin and March, 1989). When $C = 0$, the horizontal kinetic energy has zero contribution, and thus only the vertical kinetic energy is shown (black dotted line).